\documentclass[twoside]{dis09}
\usepackage[latin1]{inputenc}
\usepackage[dvips]{graphicx,epsfig,color}
\usepackage{wrapfig,rotating}
\usepackage{amssymb,amsmath,array}

\begin{document}
\title{Azimuthal decorrelations of dijets in QCD}

\author{Andrea Banfi$^1$
%
%
\vspace{.3cm}\\
%
1- Universit\`a di Milano-Bicocca and INFN, Sezione di Milano-Bicocca \\
Piazza della Scienza 3, 20126 Milano - Italy
}

\maketitle

\begin{abstract}
  We report on the status of the QCD analysis of dijet azimuthal
  decorrelations. We emphasise the relevance of resummation of soft
  and collinear enhancements in describing these observables in the
  region where the two jets are nearly back-to-back in the transverse
  plane. We also discuss the sources of theoretical uncertainties and
  possible research directions aimed at their reduction.
\end{abstract}

\section{Dijet azimuthal decorrelations}

The study of jet cross sections constitutes one of the most important
investigation tools for the physics of hadronic final states. These
observables are favoured in many aspects, both theoretically and
experimentally. First they are infrared and collinear
safe,\footnote{Some infrared unsafe cone algorithms are still on the
  market, but efforts are being made to move to infrared safe jet
  definitions~\cite{siscone}.}  which makes it possible to compute
them at all orders in QCD.  Hadronisation corrections are quite small,
being suppressed by inverse powers of the transverse energy of the
jets, and their dependence on the jet radius $R$ is under
control~\cite{jethad}.  Methods to eliminate the contamination of
underlying event and pile-up are also being developed~\cite{jetareas}.
Furthermore, measurements involving jets show less experimental
uncertainties with respect to the corresponding ones involving
particles (see for instance Ref.~\cite{hhevs-exp} for a recent
analysis).

Among jet observables we consider the azimuthal decorrelation of a
pair of jets~\cite{jetD0,jetHERA}. At tree level two jets produced in a
hard collision are highly correlated, their transverse momenta with
respect to the beam being exactly back-to-back, which implies that
their relative azimuthal angle $\Delta \phi$ is equal to $\pi$.
Additional QCD radiation decorrelates the dijet system, moving
$\Delta\phi$ away from its Born value. This feature, together with the
fact that any measurement involving angles, i.e.~ratios of momenta, is
less sensitive to jet energy scale uncertainties, make azimuthal
decorrelations an ideal ground to explore all-order properties of QCD
dynamics. In particular, in the region where $\Delta\phi$ is close to
$\pi$, we expect the dominance of configurations with multiple soft
and collinear emissions. On the contrary, when $\Delta\phi$ is small,
if the available rapidity range is large enough, one may expect the
occurrence of multiple hard emissions decorrelating the two jets.
Besides these perturbative features, azimuthal decorrelations are also
affected by intrinsic transverse momentum of partons inside the
incoming hadrons, thus making it possible to explore this
non-perturbative aspect of QCD.

As far as the experimental situation is concerned, dijet azimuthal
decorrelations have been measured both at the Tevatron~\cite{jetD0}
and at HERA~\cite{jetHERA}.  In hadronic collisions, one observes that
next-to-leading order (NLO) QCD predictions obtained with
NLOJET++~\cite{nlojet} fail for $\Delta\phi$ near $\pi$.  This region
is instead described quite well by the Monte Carlo event generators
HERWIG~\cite{herwig} and PYTHIA~\cite{pythia}. This reveals that at
the Tevatron, where the momentum fraction $x$ of struck partons is
quite large, multiple soft and collinear emissions are able to account
for the azimuthal decorrelations in the quasi-back-to-back-region.
Furthermore, due to the sensitivity of Monte Carlo predictions to
changes in the shower evolution parameters, these observables should
be exploited for the tuning of event generators~\cite{jetD0}. HERA
instead~\cite{jetHERA} explored the region of small $x$, where large
deviations from Monte Carlo predictions have been observed. However,
one finds better agreement with NLO, except of course in the region
$\Delta\phi\simeq\pi$. A question then arises on whether the Monte
Carlo tuning performed at the Tevatron at large $x$ will be able to
describe LHC data which involve smaller values of $x$.
An attempt to predict azimuthal decorrelations both at large and small
$\Delta\phi$ is represented by CASCADE~\cite{cascade}, an event
generator producing hard QCD emissions according to the CCFM
equation~\cite{ccfm}.  CASCADE, as expected, describes the azimuthal
decorrelation in the region of small $\Delta\phi$, while fixing the
distribution at $\Delta\phi$ close to $\pi$ by means of an
unintegrated parton distribution containing information on the
transverse momentum of the incoming parton inside the proton.

We decided to take another point of view. First we observe that a
resummation of soft and collinear emissions gives rise to a
distribution that rises to a constant value for
$\Delta\phi\!\to\!\pi$, consistent with what is seen in the
data~\cite{jetD0,jetHERA}.  Moreover, the agreement of NLO QCD with
data in the region of small $\Delta\phi$ suggests that in the $x$
region probed at HERA only few extra hard emissions could be
considered.  It is then reasonable to investigate whether HERA data
can be described by a resummation of soft/collinear enhancements
matched to exact NLO.

\section{Resummation of soft and collinear logarithms}

The resummation of soft and collinear logarithms is performed by
considering the observable $\Sigma(\Delta)$, the probability that
$|\pi-\Delta\phi|<\Delta$.  The azimuthal decorrelation can be then
obtained by differentiating $\Sigma(\Delta)$. The rate
$\Sigma(\Delta)$ contains logarithmic contributions up to $\alpha_s^n
\ln^{2n}\Delta$, which become large in the back-to-back region $\Delta
\ll 1$, and have to be resummed at all orders.
The first step to perform such a resummation is to investigate the
behaviour of $|\pi-\Delta\phi|$ after a single soft emission $k$,
collinear to each of the hard legs (one or two incoming and two
outgoing). If $k$ is collinear to any of the incoming partons, we have
\begin{equation*}
  |\pi-\Delta\phi| \simeq \frac{k_t}{p_{t,1}} |\sin\phi|\,,
\end{equation*}
where $k_t$ is the emission's transverse momentum (with respect to the
beam in hadronic collisions or the virtual photon momentum in DIS),
$\phi$ its azimuthal angle, and $p_{t,1}$ is the transverse momentum
of the highest-$p_t$ jet. This result does not depend on either the
jet algorithm or the recombination scheme of particles within a jet.
If instead one considers an emission collinear to one of the outgoing
legs, in any jet algorithm, soft gluon $k$ and its parent parton will
be clustered in the same jet, and the result will depend on the
recombination scheme. In any scheme that adds three-momenta
vectorially ($E$-scheme, $P$-scheme, $E0$-scheme) the azimuth of the
jet does not change, so that $|\pi-\Delta\phi|$ is completely
unaffected by emissions inside the hard jets. Therefore the azimuthal
decorrelation turns out to be a non-global
observable~\cite{nonglobal}, its resummation becomes extremely tricky
and will not be discussed here.  If instead one uses a $p_t$-weighted
recombination scheme, as is done at HERA, we have
\begin{equation*}
  |\pi-\Delta\phi| \simeq \frac{k_t}{p_{t,1}} |\sin\phi-\phi|\,,
  \quad i\in\mathrm{jet 1}\,;\quad
  |\pi-\Delta\phi| \simeq \frac{k_t}{p_{t,1}} |\sin\phi-(\pi-\phi)|\,,
  \quad i\in\mathrm{jet 2}.
\end{equation*}
For an arbitrary set of secondary soft and collinear emissions, we
have~\cite{BDD}
\begin{equation*}
  |\pi-\Delta\phi| \simeq \left|
    \sum_{i \notin \mathrm{jets}}\frac{k_{ti}}{p_{t,1}}\sin\phi_i 
    -\sum_{i \in \mathrm{jet 1}}\frac{k_{ti}}{p_{t,1}} [\phi_i-\sin\phi_i]
    -\sum_{i \in \mathrm{jet 2}}\frac{k_{ti}}{p_{t,1}} [(\pi-\phi_i)-\sin\phi_i]
  \right|\,,
\end{equation*}
and the observable is now global.  We stress that the choice of the
$p_t$-weighted scheme is observable specific. For other observables a
different choice could be needed in order to achieve the best
theoretical accuracy.

Resummation is better performed in impact parameter space, where the
result for $\Sigma(\Delta)$ schematically reads~\cite{BDD}
\begin{equation}
\label{Eq:sigma-resum}
  \Sigma(\Delta) = 
  \frac{2}{\pi}\int_0^{\infty}\!\!\frac{db}{b}\sin(b\bar\Delta)\,
  \left[
    \prod_{a=1}^{n_\mathrm{in}} 
    \frac{f_a(\mu_F/b)}{f_a(\mu_F)} 
  \right]
  e^{-R_{\mathrm{in}}(b)} \,
  e^{-R_{\mathrm{out}}(b)}\,
  S(b)\,,
  \qquad
  \bar\Delta = \Delta\, e^{-\gamma_E}\,.
\end{equation}
Each function in the above equation accommodates all-order
real or virtual corrections in selected phase space regions. The
ratios of parton densities $f_a(\mu_F/b)/f_a(\mu_F)$ embody real
and virtual contributions up to the scale $\mu_F/b$, above which
real emissions are forbidden by the observable definition. All
remaining virtual corrections are included in the \emph{radiator}
functions $R_{\mathrm{in}}(b)$, $R_{\mathrm{out}}(b)$,
containing virtual corrections collinear to incoming and outgoing legs
respectively, and the \emph{soft} function $S(b)$, accounting for
soft large-angle gluons. The above resummation is valid within
next-to-leading logarithmic (NLL) accuracy, i.e.~accounts for all
terms $\alpha_s^n \ln^{n+1}\!b$ and $\alpha_s^n \ln^n \!b$ in the
\emph{logarithm} of the Fourier transform of $\Sigma(\Delta)$. At
first order in $\alpha_s$ the expressions for $R_{\mathrm{in}}(b)$ and $R_{\mathrm{out}}(b)$ read
\begin{equation*}
  R_{\mathrm{in}}(b) = C_{\mathrm{in}} \frac{\alpha_s}{\pi} \ln^2 b\,,
  \qquad
  R_{\mathrm{out}}(b) =
  \frac{C_{\mathrm{out}}}{3} \frac{\alpha_s}{\pi} \ln^2 b\,,
\end{equation*}
where $C_{\mathrm{in}}$ and $C_{\mathrm{out}}$ are the total colour
charges of incoming and outgoing legs respectively.\footnote{For
  instance in DIS with an incoming quark $C_{\mathrm{in}}=C_F$ and
  $C_{\mathrm{out}}=C_F\!+\!C_A$.} Note that $R_{\mathrm{in}}(b)$
is the usual radiator for $p_T$ resummation (e.g.~Drell-Yan
lepton-pair $p_T$~\cite{DYpt}) while the expression for
$R_{\mathrm{out}}(b)$ has never been encountered before and
reflects the unusual dependence of $|\pi-\Delta\phi|$ on the rapidity
and azimuth of each gluon emitted from outgoing legs. The soft
function $S(b)$ is in general a matrix in the colour space
spanned by the hard emitting partons, and depends on emitters'
four-velocities.

\section{Towards phenomenology}

\begin{wrapfigure}{l}{0.5\columnwidth}
\centerline{\includegraphics[width=0.5\columnwidth,height=6cm]
  {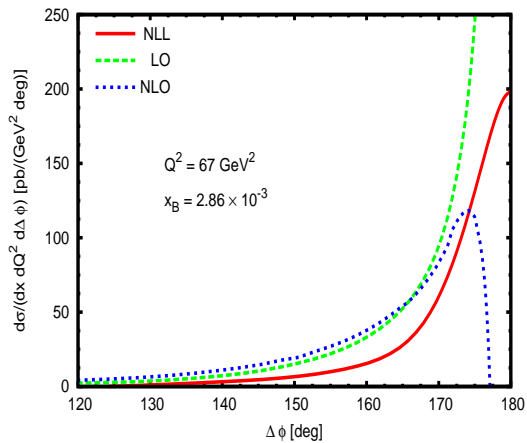}}
\caption{Fixed order and resummed predictions for azimuthal dijet
  decorrelation in DIS.}\label{Fig:fo-vs-resum}
\end{wrapfigure}

We consider azimuthal dijet decorrelation in DIS\footnote{This
  observable, defined with a $p_t$-weighted recombination scheme, is
  global. This is not so for the corresponding observable considered
  at the Tevatron, which is non-global~\cite{jetD0}. } with the same
kinematical cuts adopted by H1~\cite{jetHERA}, that is two jets with
$-1<\eta^{\mathrm{lab}} < 2.5$ and transverse momenta in the hadronic
centre-of-mass (HCM) frame $p_t \!> \! 5\mathrm{GeV}$. We give
predictions for the differential cross section in the angular distance
$\Delta\phi$ in the HCM frame.  Results are shown in
Fig.~\ref{Fig:fo-vs-resum} for a selected value of $x_B$ and $Q^2$,
reported in the figure. One can see that around $\Delta\phi\simeq
170^o$ both LO and NLO predictions obtained from NLOJET++ diverge due
to the presence of large logarithms. The NLL resummed curve in
Eq.~(\ref{Eq:sigma-resum}) instead rises to a constant value, in
qualitative agreement with what is seen in the data~\cite{jetHERA}.

\begin{wrapfigure}{r}{0.5\columnwidth}
  \centerline{\includegraphics[width=0.5\columnwidth,height=6cm]
    {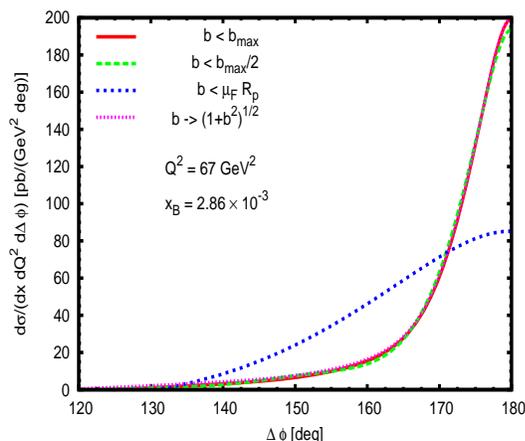}}
  \caption{The impact of different prescriptions on
    azimuthal dijet decorrelations in DIS.}\label{Fig:resum-therr}
\end{wrapfigure}

The resummed expression in Eq.~(\ref{Eq:sigma-resum}) contains some
ambiguities that have to be addressed before comparing to data. First
all functions of $b$ in the Fourier transform of $\Sigma(\Delta)$
result from integrals of the running coupling $\alpha_s(k_t)$ down to
$k_t\!=\!p_{t,1}/b$. When $b$ becomes large, this value can exceed the
Landau pole, thus making Eq.~(\ref{Eq:sigma-resum}) ill-defined.  We
then decided to cut the $b$-integral at the value $b_{\max}$
corresponding exactly to the Landau pole. However, since the
$b$-integral is strongly suppressed in the large-$b$ region, it is not
affected significantly by reducing $b_{\max}$ by a factor, as can be
seen in Fig.~\ref{Fig:resum-therr} comparing the curves labelled
$b<b_{\max}$ and $b<b_{\max}/2$. This means that even in the region
$\Delta\phi\simeq 180^o$ dynamics is not dominated by low momentum
scales. At small $b$ one has to take into account that setting $b=0$
corresponds to computing the total cross section. This imposes the
requirement that the Fourier transform of $\Sigma(\Delta)$ tend to
unity as $b\to 0$.  This is achieved by either freezing its value at
one for $b < 1$ (this is what is done in the plot in
Fig.~\ref{Fig:fo-vs-resum}) or making the replacement $b \to
\sqrt{1+b^2}$~\cite{DYpt}.  Figure~\ref{Fig:resum-therr} shows that
either choice (corresponding to the curves $b<b_{\max}$ and
$b\!\to\!(1+b^2)^{1/2}$) produces almost indistinguishable results.

The most important source of theoretical errors is the treatment of
$f_a(\mu_F/b)$, since the actual factorisation scale $\mu_F/b$ can be
smaller than the inverse size of the proton $R_p^{-1}$. The curve in
Fig.~\ref{Fig:fo-vs-resum} has been obtained by freezing the parton
densities for $b > \mu_F R_p$. However, restricting the $b$ integral
to $b < \mu_F R_p$ (see Fig.~\ref{Fig:resum-therr}) gives a result
that can be 50\% smaller in the large $\Delta\phi$ region. This
suggests that to provide accurate predictions for azimuthal
decorrelations in DIS one may need to consider the effect of the
intrinsic transverse momentum of partons inside the proton.

Finally, in order to obtain quantitative predictions, one has to match
NLL resummation to exact fixed order calculations. After matching,
$\Sigma(\Delta)$ gets multiplied by a coefficient function
$(1+C_1\alpha_s+\dots)$ in such a way that in its expansion all terms
$\alpha_s^n \ln^{2n-2}\!\Delta$ are correctly taken into account.  The
determination of $C_1$ poses two theoretical problems. The first is
that, in order to achieve such an accuracy, one has to be able to
compute $C_1$ for each hard colour configuration. This can be done in
the soft/collinear limit via a flavour jet-algorithm~\cite{jetflav},
but requires that one modify NLOJET++ to include the information on
the flavour of produced partons, as has been done in hadron-hadron
collisions in Ref.~\cite{bjets}. This work is currently in progress.
A second issue is the fact that, after subtracting all logarithms from
the LO prediction, the result tends to a constant value which is large
and negative, thus potentially giving a negative coefficient function.
This reveals that probably part of the coefficient function will have
to be resummed and eventually exponentiated, as has been done for the
Drell-Yan total cross section~\cite{ELM}.  Potentially large terms may
contain $\pi^2$, coming either from Coulomb phases or from azimuthal
integrations, or large logarithms of Bjorken $x$. The need for better
understanding of the coefficient function is one of the reasons why we
have decided to study these sub-leading contributions in the simpler
case of the $Z$ boson $a_T$ distribution in hadron-hadron
collisions~\cite{at}.  The resummation for this observable is very
similar to that for azimuthal decorrelations, with the simplification
that it does not involve any outgoing jets and that it should have no
contributions that are logarithmically enhanced by kinematics (e.g.~by
$\ln 1/x$).  These last contributions could be addressed by extending
soft collinear resummation to include also multiple hard gluon
emissions as is done for instance in CASCADE.

\section*{Acknowledgements}
It has been a pleasure to work on this subject together with Mrinal
Dasgupta and Yazid Delenda. I am also grateful to the organisers for
the cheerful and stimulating atmosphere I have experienced during the
whole workshop.


\begin{footnotesize}

\end{footnotesize}


\end{document}